\documentclass{article}
\usepackage{hiph-art}
\volnumber{19} \issuenumber{1} \edyear{2004}                             %|
\frompage{000} \topage{000}                                              %|
\recrevdate{15 May 2004}                                                 %|
\def\rightmark{The 3rd flow component}
\def\leftmark{L.P. Csernai, et al.}
 
\title{The 3rd Flow Component as a QGP Signal} 
\authors{ 
{L.P. Csernai$^{1,2}$, A. Anderlik$^1$, Cs. Anderlik$^1$, V.K. Magas$^3$,\\
  E. Moln\'ar$^1$, A. Ny\'{\i}ri$^1$, D. R\"ohrich$^1$ and K. Tamosiunas$^1$ %
\index{Csernai, L.P.}   
\index{Anderlik, A.}
\index{Anderlik, Cs.}
\index{Magas, V.K.}
\index{Moln\'ar, E.}
\index{Ny\'{\i}ri, \'A.}
\index{R\"ohrich, D.}
\index{Tamosiunas, K.}
}\\[2.812mm]
{\normalsize
\hspace*{-8pt}$^1$ Section for Theoretical and Computational Physics, University of Bergen\\ 
N-5007 Bergen, Norway\\[0.2ex] 
\hspace*{-8pt}$^2$ MTA-KFKI, Research Institute of Particle and Nuclear Physics\\ 
H-1525 Budapest, Hungary\\[0.2ex]
\hspace*{-8pt}$^3$ Physics Department, University of valencia, Dr. Moliner 50, 46100 Burjassot, Valencia, Spain
}}
 
\abstract{
Earlier fluid dynamical calculations with QGP 
show a softening of the directed flow while with hadronic matter this
effect is absent. On the other hand, we indicated that a third flow
component shows up in the reaction plane as an enhanced emission, which 
is orthogonal to the directed flow. This is not shadowed by the 
deflected projectile and target, and shows up at measurable rapidities,
$y_cm = 1-2$. To study the formation of this effect initial stages of 
relativistic heavy ion collisions are studied. An effective string 
rope model is presented for heavy ion collisions at RHIC energies. 
Our model takes into account baryon recoil for both target and projectile, 
arising from the acceleration of partons in an effective field. The typical
field strength (string tension) for RHIC energies is about 5-12 GeV/fm, 
what allows us to talk about "string ropes". The results show that  
QGP forms a tilted disk, such that the direction of the largest pressure 
gradient stays in the reaction plane, but deviates from both the beam 
and the usual transverse flow directions. The produced initial state 
can be used as an initial condition for further hydrodynamical calculations. 
Such initial conditions lead to the creation of third flow component.
Recent $v_1$ measurements are promising that this effect can be used
as a diagnostic tool of the QGP.
}

 
\makeindex
\begin{document}
 
\maketitle

\section{Introduction}\label{intro}

In high energy heavy ion reactions we observe many thousand particles
produced in the reaction, so, we have all reasons to assume that in a
good part of the reaction the conditions of the local equilibrium and 
continuum like behavior are satisfied. The initial and 
final stages are, on the other hand, obviously not in
statistically equilibrated states, and must be described separately,
in other theoretical approaches. The different approaches, as they
describe different space-time (ST) domains and the corresponding
approaches can be matched to each other across ST hyper-surfaces or
across some transitional layers or fronts. The choice of realistic
models at each stage of the collision, as well as the correct coupling of
the different stages or calculational modules are vital for a
reliable reaction model.

The fluid dynamical (FD) model plays a special role among the 
numerous reaction models, because it can be applied only 
if the matter reaches local thermal equilibrium. If this
happens the matter can be characterized by an Equation of
State (EoS). This is what we are actually looking for in these
experiments and this is why fluid dynamics is so special.

\section{The Soft Point}

The most straightforward and first observed collective flow phenomenon
was the directed transverse flow. The projectile and target
in the collision have overlapped, the participant region is
compressed and heated up, which results in high pressure, $P$.
The spectator regions, which are not compressed and heated up so
much have contact with the hot central zone. The size of the
contact surface, $S$, is of the order of the cross-section of the
nuclei or somewhat less, depending on the impact parameter.
During the collision the spectators and participants were in contact
for a period of time $\tau$ and in this time the spectators
acquired a directed transverse momentum component of
$$
p_x \approx  P \times S \times \tau
$$
At a few hundred MeV/nucleon Beam energy the directed transverse
flow momentum was almost as large as the random average transverse
momentum and it was also close to the CM beam momentum.\cite{book}

The effect was dominant, it was used to estimate the compressibility
of nuclear matter, and the general expectation was that this is a good
tool to find the threshold of the phase transition of the
Quark Gluon Plasma. The reason is that the pressure decreases considerably
versus the energy density in a first order phase transition, when the 
phase balance and formation of the new phase prevents the pressure
from increasing in the mixed phase region. All estimates indicated that
this is a strong and observable decrease in the pressure, and
it has got the name the "soft point".

To some extent the expectations were fulfilled, the directed transverse
flow decreased as the beam energy was increased above 1 GeV and further.
This was obvious compared to the beam momentum, but the decrease was
also well observable compared to the average transverse momentum.

The reason is actually simple: although at high energies the pressure
should increase the target and projectile are becoming more and more Lorentz
contracted, and the overlap time also. So, the directed transverse momentum
the non-par\-ti\-ci\-pant matter could acquire was
$$
p_x \approx  P \times \frac{S}{\gamma} \times \frac{\tau}{\gamma} \ .
$$
Thus, at high energies the directed flow could not compete with the
trivial effect of reducing the flow by $\gamma^2$. This eliminated the 
interest in the directed flow, while the elliptic flow became a dominant
and strong effect.  

\section{The Initial State}

The elliptic flow, which is observable 
in the coefficient of the 2nd harmonic, $v_2$, \cite{PV98} of 
the flow analysis, was not effected by the increasing beam energy
because it developed in the CM frame in the participant matter, due to
a special symmetry in the initial state and due to the large pressure
gradient in the direction of the reaction plane. The elliptic flow
took over the role of the directed flow, and became an important tool
in determining the basic reaction mechanism.  Only models, which
included a large collective pressure could reproduce the data, i.e.
mainly fluid dynamical models.  On the other hand, the elliptic flow
effect was so strong that most fluid dynamical models could fit the
data, even rather simple, one- and two- dimensional ones, so it was
not a very strong diagnostic tool.

A deeper insight into fluid dynamical calculations indicated that 
a similar effect, arising exclusively from participant matter, should
be possible to see also in the $v_1$ harmonic. This was first observed 
in 1999 \cite{LaDi99} based on earlier FD calculations, which included
a QGP EoS. The effect was named the 3rd flow component.

The effect was verified at the CERN SPS, but it was initially not
detected at RHIC. Now at the beginning of this year, first STAR then
two other groups have succeeded to measure the $v_1$ harmonic coefficient
of the collective flow at RHIC also. 
\cite{STAR04,Oldenburg} 
This indicated to us that one needs
extended theoretical studies, to map the sensitivity of this measurable.

As all fluid dynamical effects it depends on the initial state and on the
EoS. It is of special interest how this effect depends on the initial
state because this may shed light on the mechanisms of the formation
of QGP in heavy ion reactions.

The initial state models in
our recent works are based on a collective (or coherent)
Yang-Mills model, a versions of flux-tube models.  This approach \cite{GC86} 
was implemented in a Fire-Streak geometry streak by streak and it was
upgraded to satisfy energy, momentum and baryon charge conservations
exactly at given finite energies \cite{MCS01,MCSe02}. The effective 
string-tension was different for each streak, stretching in the
beam direction, so that central streaks with more color (and baryon) charges
at their two ends, had bigger string-tension and expanded less, than peripheral
streaks. The expansion of the streaks was {\it assumed} to last until the
expansion has stopped. Yo-yo motion, as known from the {\it Lund-model} was
not assumed. 

Our calculations show that a tilted initial state is formed, 
which leads to the 
creation of the third flow component \cite{LaDi99}, peaking at rapidities
$|y| \approx \pm 0.75$ \cite{MCSb01}. Recent, STAR $v_1$ data \cite{STAR04} 
indicate that our {\it assumption} that the
string expansion lasts until full stopping of each streak, may
also be too simple and local equilibration may be achieved earlier,
i.e. before the full uniform stopping of a streak. We did not 
explicitly calculate dissipative processes, some friction within and
among the expanding streaks is certainly present and experiments seem 
to indicate that this friction is stronger. 

The expanding strings are shown in Fig. 1

\begin{figure}[htb]
%\vspace*{-1.1cm}
                 \insertplot{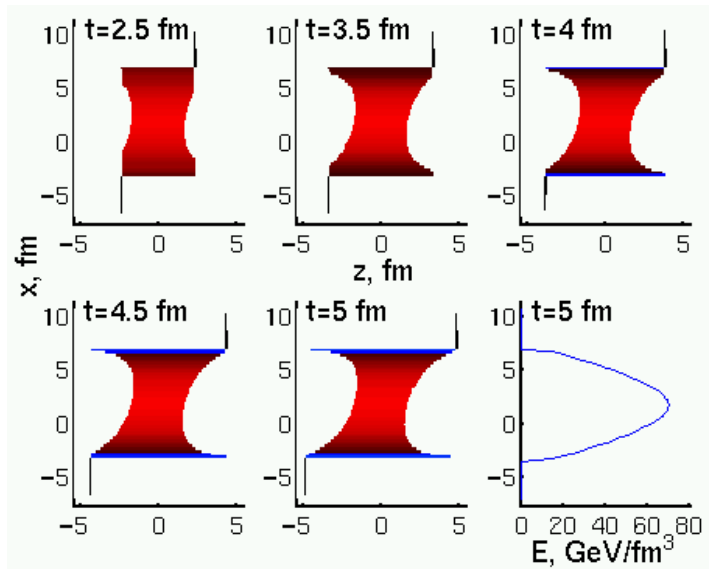}
%\vspace*{-2.1cm}
\caption[]{
Au + Au collision at $E_{cm} = 65+65$ GeV/nucl, $b = 0.25 \times 2R_{Au}$,
and the parameter determining the string tension $A = 0.08$, 
$E = T_{00}$ is presented in the reaction plane as a function 
of $x$ and $z$ for different times in the laboratory
frame. The final shape of the QGP volume is a tilted disk $\approx 45^0$, 
and the direction of the fastest expansion will
deviate from both the beam axis and the usual transverse flow direction, 
and will generate the third flow component.
Note that the streaks are moving in the CM frame. From \cite{MCSe02}
}
\label{fig1}
\end{figure}

With time the streaks expand and move to the right and left on the
top and bottom respectively. This motion will be reflected in the
subsequent FD motion also so the tilted transverse expansion will be
observed at large polar angles, i.e. relatively low rapidities.
The position of the 3rd component flow peaks depends on the
(i) impact parameter, (ii) the effective time of the left/right
longitudinal motion, (iii) the string tension determining the 
lengths of the streaks, (iv) the thickness of the configuration which
determines the initial pressure gradient. The string tension varies
as a function of the distance of a given streak from the central beam
axis, because of the amount of the matter at the ends of the streaks.
This determined the amount of color charge, and so the string tension
is
$$
\sigma = A \left( \frac{E_{cm}}{m} \right)
 \frac{\sqrt{N_1 N_2}}{\delta x  \ \delta y}
$$
given in terms of the Baryon charges at the two ends of the streak, $N_1$ and $N_2$,
and the cross section of the string, $ \delta x\  \delta y$.

Although, there are several effects determining the angle of the 3rd flow 
component, some of these can be traced down by other measurements also:
the impact parameter by the multiplicity, the effective expansion time 
and size by two particle correlation measurements, etc. Thus, there is a
reasonable possibility that the $v_1$ measurements will lead to a more detailed 
insight to the details of the QGP formation in heavy ion reactions. 

The subsequent fluid dynamical calculations show the development of the
3rd flow component from these type of initial states. These FD model calculations
have to be supplemented with a Freeze Out model. There is a very essential
development in this field also, and we can have different levels of approach
from simplified freeze out descriptions across an FO hypersurface using
an improved Cooper-Fry description, or a full kinetic description, originating
from a modified Boltzmann Transport Equation approach. This final step
is complicated further by the fact that the data indicate a rapid and simultaneous 
hadronization and freeze out. 

Nevertheless, we expect that the $v_1$ flow data will be sufficiently detailed
and accurate soon, and the 3rd flow component will be an effect which is
sufficiently strong that the complex final effects at the freeze out
will allow a successful analysis of the matter properties via these flow
measurements.
 
\section*{Acknowledgments}
One of the authors,  L.P. Cs., thanks the Alexander von Humboldt Foundation
for extended support in continuation of his earlier Research Award.
The authors thank the hospitality of the Frankfurt Institute of Advanced
Studies and the Institute for Theoretical Physics of the University of 
Frankfurt, the Gesellschaft f\"ur Schwerionenforschung, and the University
of Giessen where parts of this work were done.
 
%\section*{Notes}
%\begin{notes}
%\item[a]
%\end{notes}

\vfill\eject
\end{document}